\documentclass[10pt,twocolumn,a4paper]{article}
\usepackage{subcaption}
\usepackage[margin=2cm]{geometry}
\usepackage{times}
\usepackage{amsmath,amssymb}
\usepackage{graphicx}
\usepackage{booktabs}
\usepackage{url}
\usepackage{titlesec}

\titleformat{\section}
  {\Large\bfseries\centering}{\thesection.}{0.5em}{}

\titleformat{\subsection}
  {\normalsize\bfseries}{\thesubsection}{0.5em}{}

\renewenvironment{abstract}{
  \section*{Abstract}
}{\par\vspace{1em}}

\title{LG Uplus System with Multi-Speaker IDs and Discriminator-based Sub-Judges for the WildSpoof Challenge}

\author{Jinyoung Park$^{1}$, Won Jang$^{1}$, Jiwoong Park$^{1}$ \\
{\small $^{1}$LG Uplus, South Korea}\\
{\small \texttt{\{wlsdud7907,taylor,jivvp\}@lguplus.co.kr}}
}

\date{}

\begin{document}
\maketitle

\begin{abstract}
This paper describes our submission to the WildSpoof Challenge Track~2, which focuses on spoof-aware speaker verification (SASV) in the presence of high-quality text-to-speech (TTS) attacks. We adopt a ResNet-221 backbone and study two speaker-labeling strategies, namely Dual-Speaker IDs and Multi-Speaker IDs, to explicitly enlarge the margin between bona fide and generated speech in the embedding space. In addition, we propose discriminator-based \emph{sub-judge} systems that reuse internal features from HiFi-GAN and BigVGAN discriminators, aggregated via multi-query multi-head attentive statistics pooling(MQMHA). Experimental results on the SpoofCeleb corpus show that our system design is effective in improving agnostic detection cost function (a-DCF).
\end{abstract}

\section{Introduction}
The WildSpoof Challenge provides a realistic benchmark for evaluating spoof-aware speaker verification (SASV) systems under diverse text-to-speech (TTS) attacks and recording conditions. Track~2 focuses on speaker verification in the presence of high-quality synthetic and converted speech, where the system must discriminate both the target identity and the authenticity of the input (bona fide vs. generated).

In this work, we focused on the similarity between the objective loss of TTS discriminators and the spoofed voice detection task. We suggest a auxiliary classifiers for spoof detection that reuses the discriminator architectures from HiFi-GAN\cite{kong2020hifigan} and BigVGAN\cite{lee2023bigvgan}. Building a robust SASV system for this track requires (i) strong speaker-discriminative representations and (ii) fine-grained cues that can distinguish human recordings from high-quality neural TTS.

To this end, we start from a ResNet-221 backbone that we have frequently used for spoofing aware speaker verification tasks. On top of this backbone, we explore two ways of assigning speaker labels that explicitly encode the relationship between bona fide speech and its corresponding generated attacks. Furthermore, inspired by the success of adversarial training in neural TTS, we reuse discriminator architectures from popular vocoders as auxiliary \emph{sub-judges} to provide additional spoof-discriminative information.

\section{Proposed System}
\begin{figure}[t]
  \centering
  \begin{subfigure}[t]{\linewidth}
    \centering
    \includegraphics[width=\linewidth]{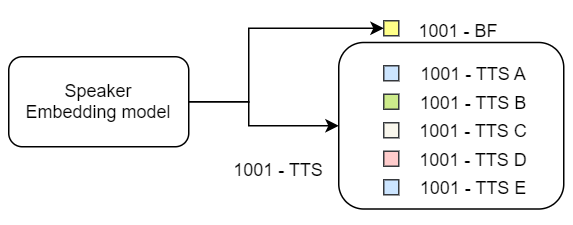}
    \caption{Dual-Speaker IDs.}
    \label{fig:dual}
  \end{subfigure}

  \vspace{0.5em}

  \begin{subfigure}[t]{\linewidth}
    \centering
    \includegraphics[width=\linewidth]{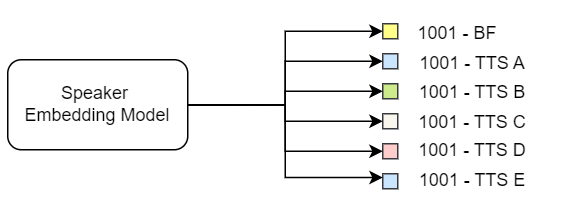}
    \caption{Multi-Speaker IDs.}
    \label{fig:multi}
  \end{subfigure}

  \caption{Illustration of the two label-design strategies. \emph{BF} defines one identity that contains bona fide speech of the speaker. 
    In Dual-Speaker IDs, TTS A--E is grouped into another identity that contains speech from multiple TTS models. 
    In Multi-Speaker IDs, TTS A--E are each assigned a different speaker identity and represent utterances generated by distinct TTS models.}
  \label{fig:ids}
\end{figure}

\subsection{Backbone SASV System by Dual/Multi-Speaker IDs}
Our main SASV backbone is based on a ResNet-221 convolutional architecture trained on log-mel spectrograms. The final frame-level features are aggregated by a statistics pooling layer to produce a fixed-dimensional embedding, followed by a classification head trained with a margin-based softmax loss. The goal is to obtain embeddings where bona fide and generated speech are well separated while preserving speaker discriminability.

We investigate two label-design strategies, illustrated in Figure~1 and Figure~2.

\paragraph{Dual-Speaker IDs (Figure 1).}
For each real speaker (e.g., \texttt{spk1001}), we define \emph{two} speaker IDs for the bona fide speaker and virtual speaker for all generated utterances (TTS, VC, etc.).
Thus, bona fide and generated samples associated with the same target speaker are forced to occupy different regions in the embedding space, with an explicit angular margin between the two virtual speakers. This encourages the model to learn robust cues that distinguish real from synthetic speech while remaining aware of the underlying target identity.

\paragraph{Multi-Speaker IDs (Figure 2).}
In the second strategy, we still assign a single ID to each bona fide speaker, but we introduce additional IDs for each \emph{TTS model} (or attack type). This design treats each TTS model as a distinct speaker in the classification task. The backbone is thus encouraged to capture model-specific artifacts shared across speakers, which can be beneficial for detecting and generalizing to unseen spoofing attacks.

\subsection{Discriminator-based Sub-Judge System}
Modern neural TTS systems typically employ discriminators during training to encourage the generated waveform to be perceptually similar to natural speech. Inspired by the similarity between the objective loss of TTS discriminators and the spoofed voice detection task, we propose a \emph{sub-judge} system that reuses off-the-shelf discriminator architectures from HiFi-GAN\cite{kong2020hifigan} and BigVGAN\cite{lee2023bigvgan} as auxiliary classifiers for spoof detection on SpoofCeleb. We extracted all possible layers, which HiFi-GAN employs a 5 layers from multi-period discriminator (MPD) and 3 layers multi-scale discriminator (MSD). On the other hand, BigVGAN uses 5 layers from multi-period discriminator (MPD) as well, but replaces MSD with 3 layers of multi-resolution spectrogram discriminator (MRD).
To integrate such features into a single fixed-dimensional representation, we adopt multi-query multi-head attentive statistics pooling (MQMHA)\cite{chung2021mqmha}. We concatenated results from MQMHA pooling layers from each sub-discriminators. This yields a unified embedding that reflects multi-resolution discriminator judgments.

\paragraph{Classification head and training.}
The Binary spoofing label (bona fide vs. generated) is used for the classifcation head layer training. We implemented with few fully-connected layer and trained the sub-judge systems on the SpoofCeleb training set using a standard cross-entropy loss.

\section{Experiments}

\begin{table}[t]
    \centering
    \caption{Results of different speaker-ID label on SpoofCeleb dev set.}
    \label{tab:backbone}
    \begin{tabular}{lcc}
        \toprule
        System & EER (\%) & min-DCF \\
        \midrule
        Dual-Speaker IDs & 14.666 & 0.896 \\
        Multi-Speaker IDs & 7.352 & 0.683 \\
        \bottomrule
    \end{tabular}
\end{table}

\begin{table}[t]
    \centering
    \caption{Effect of discriminator-based sub-judges on SASV performance. EER and a-DCF\cite{a-dcf} values are left blank for later completion.}
    \label{tab:subjudge}
    \begin{tabular}{lcc}
        \toprule
        System & EER (\%) & a-DCF \\
        \midrule
        Original (no sub-judge) & 8.464 & 0.1422 \\
        + HiFi-GAN only & 8.604 & 0.1383 \\
        + BigVGAN only & 8.398 & 0.1415 \\
        + HiFi-GAN \& BigVGAN & 8.521 & 0.1363 \\
        \bottomrule
    \end{tabular}
\end{table}

We used SpoofCeleb train set to train our model and dev set to evaluate our model. ResNet-221 backbone is trained with log-mel spectrograms extracted from 16~kHz waveforms. We compare the Dual-Speaker ID and Multi-Speaker ID strategies described in Section~2.1 under the same training recipe.

For the sub-judge experiments, we train separate HiFi-GAN-based and BigVGAN-based sub-judges on the same training data. At evaluation time, we consider four configurations: (i) the original backbone only, (ii) backbone + HiFi-GAN sub-judge, (iii) backbone + BigVGAN sub-judge, and (iv) backbone + both sub-judges. Performance is reported in terms of equal error rate (EER) and agnostic detection cost function (a-DCF)\cite{a-dcf} on the development and evaluation subsets.

Table~\ref{tab:backbone} summarizes the effect of different speaker-labeling strategies on SpoofCeleb dev set. The results suggest that explicitly modeling generated speech either as a per-model identities can improve the separation between bona fide and spoofed samples in the embedding space.
Table~\ref{tab:subjudge} shows the impact of adding the discriminator-based sub-judges. The HiFi-GAN and BigVGAN sub-judges provide complementary cues, and combining both tends to yield further improvements in EER and a-DCF compared to the original backbone. Since the sub-judges are specifically optimized to detect spoofed voices, they may slightly increase the overall EER by introducing additional false alarms on bona fide trials. However, under the WildSpoof challenge metric a-DCF, which assigns a higher cost to missed detections of spoofed attacks than to false alarms, the sub-judges lead to a substantial reduction in cost, demonstrating their effectiveness in mitigating the practically more critical spoofed-voice attacks.

\section{Conclusion}
In this paper, we presented our SASV system for the WildSpoof Challenge Track~2, built on a ResNet-221 backbone with speaker-ID relabeling and discriminator-based sub-judges. We found that the Multi-Speaker ID strategy consistently outperformed the Dual-Speaker ID strategy in terms of both EER and a-DCF, indicating that explicitly modeling each TTS model as a separate speaker is beneficial for spoof-aware verification. In addition, incorporating discriminator-based sub-judges derived from HiFi-GAN and BigVGAN further improved performance on a-DCF, improving performance on detecting spoof cases.

\end{document}